\documentclass[11pt]{amsart}   
\usepackage{amssymb,xy}
\usepackage{amsthm, amsmath, epsfig}
\usepackage[latin1]{inputenc}
\usepackage[T1]{fontenc}
\xyoption{all}
\newcommand\x{{\bf x}}

\newcommand\bfa{{\bf a}}

\newcommand\bff{{\bf f}}

\newcommand\zero{{\bf 0}}
\newcommand\cc{{\mathbb C}}
\newcommand\zz{{\mathbb Z}}

\theoremstyle{plain}

\newtheorem{theorem}{Theorem}[section]

\newtheorem{proposition}[theorem]{Proposition}

\theoremstyle{definition}

\newtheorem{definition}[theorem]{Definition}
\newtheorem{example}[theorem]{Example}
\newtheorem{algorithm}[theorem]{Algorithm}

\theoremstyle{remark}

\title[Computing all Affine Solution Sets of Binomial Systems]
{Computing all Affine Solution Sets
 of Binomial Systems (extended abstract)}
\author{
Danko Adrovic
\and
Jan Verschelde}
\date{29 April 2014}

\address{\small \rm  University of Illinois at Chicago,
Department of Mathematics, Statistics, and Computer Science,
851 S. Morgan Street (m/c 249), Chicago, IL 60607-7045, USA}
\email{\{adrovic,jan\}@math.uic.edu (Corresponding author: Jan Verschelde)}

\begin{document}

\begin{abstract}
To compute solutions of sparse polynomial systems efficiently
we have to exploit the structure of their Newton polytopes.
While the application of polyhedral methods naturally excludes
solutions with zero components, an irreducible decomposition of
a variety is typically understood in affine space,
including also those components with zero coordinates.
For the problem of computing solution sets in the intersection of
some coordinate planes, the direct application of a polyhedral method fails,
because the original facial structure of the Newton polytopes may alter
completely when selected variables become zero.
Our new proposed method enumerates all factors contributing to a generalized
permanent and toric solutions as a special case of this enumeration.
For benchmark problems such as the adjacent 2-by-2 minors of a general
matrix, our methods scale much better than the witness set representations
of numerical algebraic geometry.
\end{abstract}

\maketitle

\section{Introduction}
Our investigation in~\cite{AV13b}
starts with the sparsest kind of polynomial systems:
those with exactly two monomials with nonzero coefficients in every equation.
This sparsest type of systems is called {\em binomial}.
Software implementations of primary decompositions 
of binomial ideals~\cite{ES96} are described in~\cite{BLR99} and~\cite{OP00}.
Recent algebraic algorithms are developed in~\cite{Kah10} and~\cite{Oje11}.
The complexity of counting the total number of
affine solutions of a system of $n$ binomials
in~$n$ variables was shown as \#P-complete~\cite{CD07}.
In \cite{HJS12} combinatorial conditions for the existence of 
positive dimensional solution sets are given,
for use in a geometric resolution.
%
Symbolic polyhedral algorithms for 
computing isolated roots of sparse systems are in~\cite{HJS10}.


\section{Monomial Maps representing Affine Solution Sets}

Solution sets of binomial systems can be described as monomial maps,
obtained via unimodular coordinate transformations~\cite{AV12},
see also~\cite{GW12} and~\cite{HL12}.
Note that some sparse polynomial systems such as the
cyclic~$n$-roots problems have monomial maps as solution sets~\cite{AV13a}.

\begin{definition} \label{defmonpar}
{\rm A {\em monomial map} of 
a $d$-dimensional solution set in~$\cc^n$ is
\begin{equation} \label{eqmonpar}
   x_k = c_k t_1^{v_{1,k}} t_2^{v_{2,k}} \cdots t_d^{v_{d,k}},
   \quad c_k \in \cc, v_{i,k} \in \zz,
\end{equation}
for $i=1,2,\ldots,d$ and $k=1,2,\ldots,n$.
}
\end{definition}
For a toric solution, all coefficients $c_k$ in the monomial 
map~(\ref{eqmonpar}) are nonzero.  
For an affine solution set, several coordinates may be zero.  
When setting variables to zero, it may happen that all constraints on 
some other variables vanish, then we say that those variables are {\em free},
while others are still linked to a toric solution of a subset 
of the original equations.

\section{A Generalized Permanent}

To enumerate all choices of variables to be set to zero,
we use the matrix of exponents of the monomials to define
a bipartite graph between monomials and variables.

\begin{definition}  {\rm Let $\bff(\x) = \zero$ be a system.
We collect all monomials $\x^\bfa$ that occur in $\bff$ along
the rows of the matrix, yielding the {\em incidence matrix}
\begin{equation}
   M_\bff[\x^\bfa,x_k] = 
   \left\{
      \begin{array}{lcl}
         1 & {\rm if} & a_k > 0 \\
         0 & {\rm if} & a_k = 0. \\
      \end{array}
   \right.
\end{equation}
Variables which occur anywhere with a negative exponent
are dropped. }
\end{definition}

\begin{example} {\rm For all adjacent minors of a 2-by-3 matrix,
the incidence matrix is
\begin{equation}
   M_\bff = 
   \left[
\begin{array}{c|ccccccc}
                & x_{11} & x_{12} & x_{13} & x_{21} & x_{22} & x_{23} \\ \hline
  x_{11} x_{22} &   1    &   0    &   0    &   0    &   1    &   0    \\
  x_{21} x_{12} &   0    &   1    &   0    &   1    &   0    &   0    \\
  x_{12} x_{23} &   0    &   1    &   0    &   0    &   0    &   1    \\
  x_{22} x_{13} &   0    &   0    &   1    &   0    &   1    &   0    \\
\end{array}
  \right]
\end{equation}
for the system defined by $\bff = (f_1, f_2)$
with $f_1 = x_{11} x_{22} - x_{21} x_{12}$
and $f_2 = x_{12} x_{23} - x_{22} x_{13}$.
For this example, the rows of $M_\bff$ equal the exponents of the monomials.
We select $x_{12}$ and $x_{22}$ as variables to be set to zero,
as overlapping columns $x_{12}$ with $x_{22}$ gives all ones. }
\end{example}

\begin{proposition}
Let $S$ be a subset of variables such that
for all $\x^\bfa$ occurring in $\bff(\x) = \zero$: $M[\x^\bfa,x_k] = 1$, 
for $x_k \in S$, then setting all $x_k \in S$ to zero
makes all polynomials of $\bff$ vanish.
\end{proposition}
 
\noindent {\em Proof.} 
$M[\x^\bfa,x_k] = 1$ means: $x_k = 0 \Rightarrow \x^\bfa = 0$.
If the selection of the variables in~$S$ is such that all monomials
in the system have at least one variable appearing with positive power,
then setting all variables in~$S$ to zero makes all monomials in the
system vanish.~\qed

Enumerating all subsets of variables so that $\bff$ vanishes 
when all variables in a subset are set to zero is 
similar to a row expansion algorithm on~$M_\bff$ for a permanent:

\begin{algorithm}[recursive subset enumeration via row expansion of permanent]
\label{algenumerate} {\rm

\begin{tabbing} \\
\hspace{1cm} \= Input: \= $M_\bff$ is the incidence matrix 
                          of $\bff(\x) = \zero$; \\
             \>        \> index of the current row in $M_\bff$; and \\
             \>        \> $S$ is the current selection of variables. \\
             \> Output: all $S$ that make the entire $\bff$ vanish. \\
 \> if \= $M[\x^\bfa,x_k] = 1$ for some $x_k \in S$ \\
 \>    \> then \= print $S$ if $\x^\bfa$ is at the last row of $M_\bff$ 
                            or else go to the next row \\
 \>    \> else \> for \= all $k$: $M[\x^\bfa,x_k] = 1$ do \\
 \>    \>      \>     \> $S := S \cup \{ x_k \}$ \\
 \>    \>      \>     \> if \= $\x^\bfa$ is at the last row of $M_\bff$ \\
 \>    \>      \>     \>    \> then print $S$ \\
 \>    \>      \>     \>    \> else go to the next row \\
 \>    \>      \>     \> $S := S \setminus \{ x_k \}$
\end{tabbing}
}
\end{algorithm}

Greedy enumeration strategies can be applied in the algorithm above.
The enumeration may generate subsets of variables that lead to
affine monomial maps that are contained in other solution maps.
For detailed membership tests we refer to~\cite{AV13b}.

\newpage
\section{Computational Experiments} 

The polynomial equations of adjacent minors
are defined in~\cite[page~631]{HS00}:
$x_{i,j} x_{i+1,j+1} - x_{i+1,j} x_{i,j+1} = 0$,
$i=1,2,\ldots,m-1$, $j = 1,2,\ldots,n-1$.
For $m=2$, the solution set is pure dimensional of degree~$2^n$ and
of dimension $2 n - (n-1) = n+1$, the number of irreducible
components of $X$ equals the $n$th Fibonacci 
number~\cite[Theorem~5.9]{Stu02}.

For a pure dimensional set, we restrict the enumeration:
for every variable we set to zero, one equation has to vanish as well.
Table~\ref{tabadmbarplot} 
shows the comparison with a witness set construction, 
computed with version 2.3.70 of PHCpack~\cite{Ver99}.
Note that our method returns the irreducible decomposition,
which is more than just a witness set.
This system is one of the benchmarks in~\cite{BDHPPSSW14}, 
but neither Bertini~\cite{BHSW06} nor Singular~\cite{DGPS11} 
can get as far as our method.

\begin{table}[hbt]
{\small
\begin{tabular}{r|r|r|r|rr}
 $n$ & $2^{n-1}$ & \#maps & search & witness & \hspace{8cm} \\ \cline{1-5}
  3  &        4  &      2 &   0.00 &   0.03 \\
  4  &        8  &      3 &   0.00 &   0.16 \\
  5  &       16  &      5 &   0.00 &   0.68 \\
  6  &       32  &      8 &   0.00 &   2.07 \\
  7  &       64  &     13 &   0.01 &   7.68 \\ 
  8  &      128  &     21 &   0.01 &  28.10 \\
  9  &      256  &     34 &   0.02 &  71.80 \\
 10  &      512  &     55 &   0.05 & 206.01 \\
 11  &     1024  &     89 &   0.10 & 525.46 \\
 12  &     2048  &    144 &   0.24 & ---~~~ \\
 13  &     4096  &    233 &   0.57 & ---~~~ \\
 14  &     8192  &    377 &   1.39 & ---~~~ \\
 15  &    16384  &    610 &   3.33 & ---~~~ \\
 16  &    32768  &    987 &   8.57 & ---~~~ \\
 17  &    65536  &   1597 &  21.36 & ---~~~ \\
 18  &   131072  &   2584 &  55.95 & ---~~~ \\
 19  &   262144  &   4181 & 140.84 & ---~~~ \\
 20  &   524288  &   6765 & 372.62 & ---~~~ \\
 21  &  1048576  &  10946 & 994.11 & ---~~~ \\
\end{tabular}
}
\caption{The construction of a witness set for all adjacent minors
of a general 2-by-$n$ matrix requires the tracking of $2^{n-1}$ paths
and is much more expensive than the combinatorial search.
For $n$ from 3 to 21 column~3 lists times in seconds 
on one core at 3.49GHz for the 
combinatorial search and times ($<$ 1,000 seconds) 
for the witness construction are in the last columns. }
\label{tabadmbarplot}
\end{table}

\begin{picture}(400,0)(0,0)
\put(190,100){\epsfig{figure=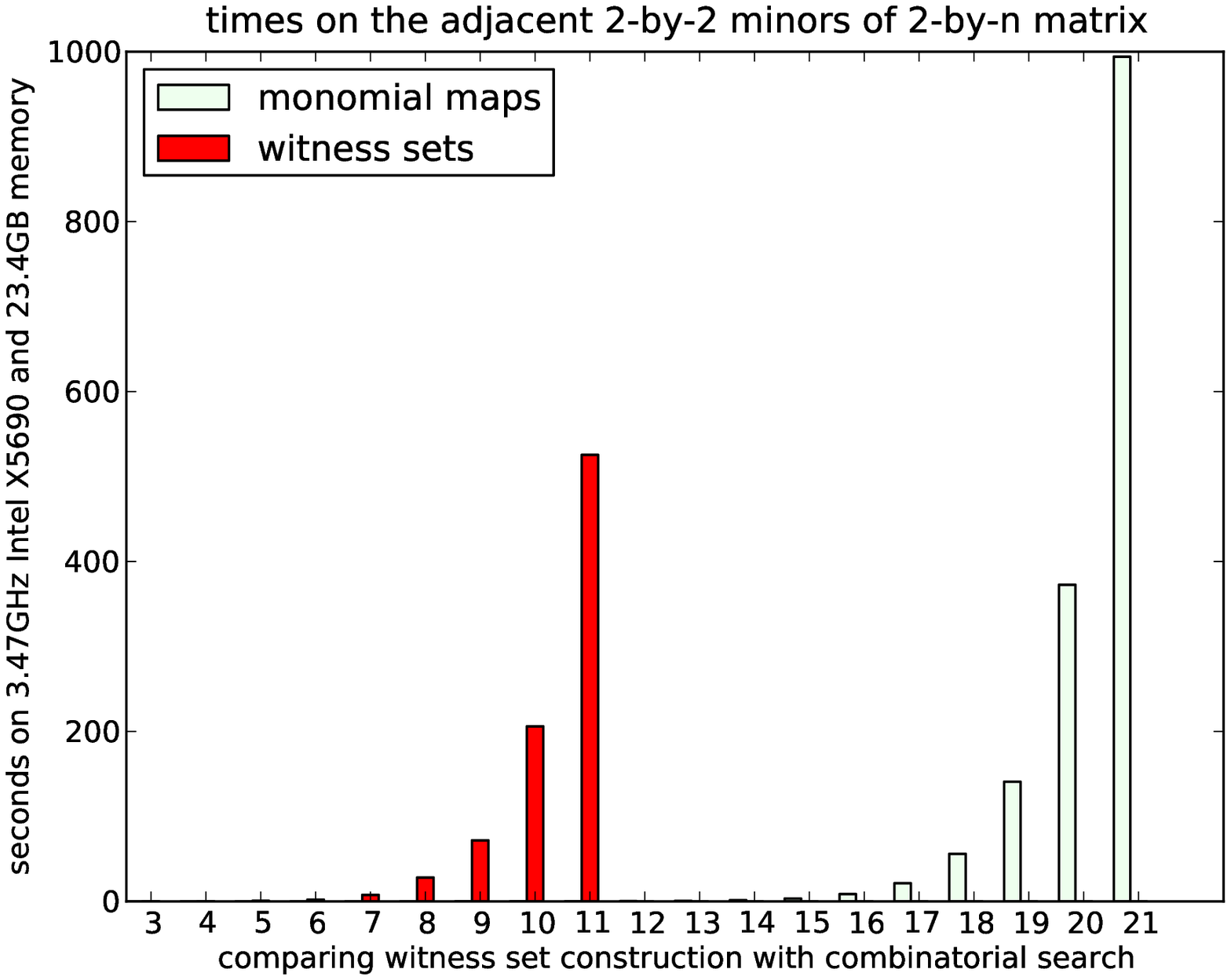,width=9cm}}
\end{picture}


Table~\ref{tabbinomials} shows timings of the 
{\tt binomialCellularDecomposition} in the
{\tt Binomials}~\cite{Kah12} package of Macaulay2~\cite{M2}
applied to the ideal defined by the adjacent minors.
\begin{table}[b!]
\begin{center}
\begin{tabular}{c|ccccccccccccc}
  $n$ &   3~ &   4~ &   5~ &   6~ &   7~ &   8~ &   9~ &  10~ 
      &  11~ &  12~ &  13~ &  14~ &  15~ \\ \hline
 time & 0.01 & 0.03 & 0.06 & 0.11 & 0.24 & 0.49 & 0.98 & 1.97
      & 4.11 & 8.96 & 22.3 & 54.7 & 160.8
\end{tabular}
\caption{CPU time in seconds on one 3.49GHz core on the adjacent minors.}
\label{tabbinomials}
\end{center}
\end{table}

As for the adjacent minors of a general 2-by-$n$
matrix the number of components returned by the cellular decomposition 
equals the number of components in an irreducible decomposition,
the comparison seems fair.

\bibliographystyle{plain}

\end{document}